\documentclass[9pt,twocolumn,twoside]{osajnl}
\usepackage[per=slash, loctolang={UK:english}, range-units=single, range-phrase=-, separate-uncertainty]{siunitx}
\journal{ol} 

\setboolean{shortarticle}{true}

\title{Optical design for laser tweezers Raman spectroscopy setups for increased sensitivity and flexible spatial detection}

\author[1]{Tobias Dahlberg}
\author[1,*]{Magnus Andersson}

\affil[1]{Department of Physics, Umeå University, 901 87 Umeå, Sweden}

\affil[*]{Corresponding author: magnus.andersson@umu.se}

\begin{abstract}
We demonstrate a method to double the collection efficiency in Laser Tweezers Raman Spectroscopy (LTRS) by collecting both the forward and back-scattered light in a single-shot multitrack measurement. Our method can collect signals at different sample volumes, granting both the pinpoint spatial selectivity of confocal Raman and the bulk sensitivity of non-confocal Raman simultaneously. Further, we display that our approach allows for reduced detector integration time and laser power. Thus, our method will enable the monitoring of biological samples sensitive to high intensities for longer times. Additionally, we demonstrate that by a simple modification, we can add polarization sensitivity and retrieve extra biochemical information.

\end{abstract}

\setboolean{displaycopyright}{true}

\begin{document}

\maketitle

\section{Introduction}
Laser tweezers Raman spectroscopy (LTRS) is a powerful technique that combines the manipulation capabilities of optical tweezers, and the biochemical fingerprinting of Raman spectroscopy \cite{Thurn:84firstLTRS,Redding_2015LTRSReview}. LTRS has been applied to study many biological systems, for example, bacteria, cells, and spores. Specifically, LTRS has been used to assess germination kinetics, spore deactivation \cite{sporegermination,sporeinactivation}, and red blood cell oxygenation states and interaction with silver nanoparticles \cite{rbcoxygenation,rbcnanoparts}. However, spontaneous Raman scattering is a weak process as only about 1 in 10 million excitation photons scatters in this manner. Combining this with the weak scattering inherent to many biological materials \cite{kuhar2018challenges}, we need either long integration times \cite{Pilat2018LongIntegration} or high excitation powers \cite{Huang2007highPower} to get clear Raman spectra. However, both of these solutions have adverse side effects. Long integration times cause the time resolution to suffer, making it impossible to investigate fast processes. Further, increasing the excitation power increases the rate of photodamage linearly, meaning that doubling the power halves the maximum measurements time  \cite{Photodamage}.

We can avoid these problems by moving away from spontaneous Raman scattering to more sensitive Raman techniques such as CARS or SERS, improving sensitivity by several orders of magnitude. However, these techniques have limitations. CARS requires expensive and complicated experimental setups to realize \cite{CARS}, and SERS necessitates the use of expensive sample substrates that often suffer problems with producing homogeneous and repeatable Raman signals \cite{SERS}. Thus, to keep the simple experimental setup of a spontaneous Raman setup and to minimize these problems, it is desirable to make it as efficient as possible. 

In this work, we show an easy way of doubling an LTRS setup's collection efficiency. We modify an LTRS setup working in the back-scattering configuration to simultaneously collect the forward-scattered light in a single-shot multitrack measurement using a high numerical aperture condenser. This concept of increasing collection efficiency in Raman spectroscopy has been explored in previous publications \cite{Anupam2009,Wu2017}. However, these approaches rely on special sample substrates to work increasing complexity compared to using standard coverslips. Thus, most LTRS setups described in the literature works by exclusively collecting either the back or forward-scattered light \cite{forwback,back1,back2}. Since LTRS samples often are transparent and the setups in general already have a high numerical aperture condenser mounted, extending the capabilities to collect both back and forward-scattered light is straightforward. Using this approach, we demonstrate a more sensitive and flexible experimental setup that allows for both confocal and non-confocal detection. Additionally, we show how to add polarization sensitivity to the setup to measure how the Raman scattering rotates polarization, which depends on the symmetry of the molecular vibrations \cite{Zhang:09SingleshotPolarize}. We anticipate that these results can be useful for future studies of sensitive biological systems.

\section{Experimental}
\begin{figure}[h]
\centering
{\includegraphics[width=\linewidth]{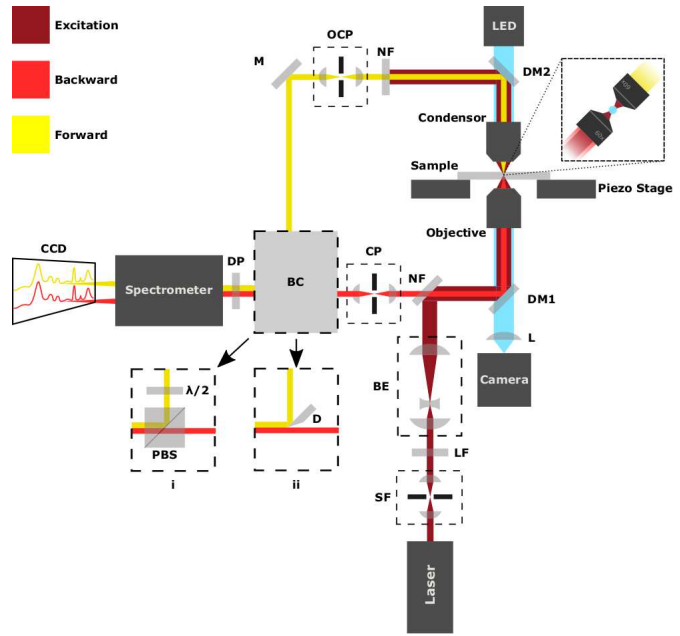}}
\caption{Illustration of the experimental setup with abbreviated component names. M mirror, DM1 750 $nm$ shortpass filter, DM2 650 $nm$ shortpass, PBS polarizing beamsplitter,OCP optional confocal pinhole, CP confocal pinhole, SF spatial filter, BE beam expansion, L lens, NF notch filter, DP depolarizer, BC beam combiner: has two options i) the two beams are merged using a PBS. A half-wave plate rotates the polarization of one beam to select between depolarized and polarized Raman scattering. ii) A pick-off mirror, D, is used to combine the beams. The inset shows a zoomed in view of the sample.}
\label{fig:system}
\setlength\belowcaptionskip{0pt}
\end{figure}

To trap particles and acquire their Raman spectra, we use a custom-built LTRS instrument built around a modified inverted microscope (IX71, Olympus) previously described in \cite{stangner2018cooke, SPIE,ANALCHEM}, see Fig. \ref{fig:system}. In short, we focus an 808 $nm$ continuous-wave laser (CRL-DL808-120-S-US-0.5, CrystaLaser) using a 60x water immersion objective (UPlanSApo60xWIR 1.2NA, Olympus) to form the trap and excite the Raman scattering. Further, we use the objective to collect the back-scattered Raman light.

For this study, we modify the setup by adding a 60x oil immersion objective (UPLANAPO60X 1.4NA, Olympus) above the sample to collect the forward-scattered light and act as a condenser to illuminate the sample. We then separate the forward-scattered light and illumination light using a shortpass filter (FESH0750, Thorlabs). After this, we pass the forward-scattered light through a notch filter, an optional confocal pinhole, and combine it with the back-scattered light using a D-shaped pick off mirror (PFD10-03-P01, Thorlabs). Next, we depolarize the beams using a depolarizer (DPP25-B, Thorlabs) and direct them to the spectrometer (Model 207, 600lp/mm grating, McPherson), where they are dispersed and imaged on separate tracks on a CCD detector (Newton 920N-BR-DD XW-RECR, Andor). Optionally, for polarization Raman measurements, we instead combine the beams using a polarizing beam splitter (PBS252, Thorlabs) with a half-wave plate inserted in the forward-scattered beam path. This change allows us to rotate the forward-scattered beam's polarization and select if we want to measure the polarized or depolarized Raman bands. With this setup, we can separately measure the forward and back-scattered light or the polarized and depolarized Raman spectrum in a single shot similar to \cite{Zhang:09SingleshotPolarize}.

To prepare a sample for use in the LTRS setup, we first construct a sample chamber. We prepared a typical sample chamber by adding two pieces of double-sided scotch tape (3M Company) spaced 5 $mm$ apart on a 24.0 × 60.0 $mm$ coverslip (no 1, Paul Marienfeld GmbH \& Co). Then, we position a 20.0 × 20.0 $mm$ coverslip (no 1 , Paul Marienfeld GmbH \& Co) on top of the tape, to form a 5.0 × 20.0 × 0.1 $mm^3$ chamber. Next, we fill the chamber with our sample liquid of choice by adding a few microliters of liquid at one of the openings and allowing capillary forces to fill the chamber. For this study, this was either 3 $\mu m$ CML polystyrene latex beads (Thermo Fisher Scientific) suspended in either absolute ethanol (VWR International) or Milli-Q water, alternatively pure cyclohexane (Honeywell International), or \textit{Bacillus Thuringiensis} (\textit{B. Thuringiensis}) spores ATCC 35646 suspended in Milli-Q water. Finally, we seal the open ends of the chamber using vacuum grease (Dow Corning).

To measure the Raman spectrum, we mount the sample to the LTRS setup and position the focal plane of the trapping objective $\sim$  30 $\mu m$ from the top coverslip. Then, we focus the condenser objective to be parfocal with the trapping objective, that is, to have overlapping focal planes. From this position, we can either measure using this parfocal setup or defocus the two objectives by moving the trapping objective up or down. After focusing, we can open the shutter to the laser and trap a particle or illuminate the sample liquid. Additionally, we can choose if we want to use the optional confocal pinhole in the forward scattered beam path for confocal or non-confocal measurements. Next, we record the Raman spectrum using 2 accumulations with 5 $s$ exposure time with a laser power of ~20 $mW$. When measuring on spores, we instead use quartz coverslips (MODEL) to construct the sample chamber and we use 2 accumulations of 30 $s$ and a laser power of 2.5 $mW$.

\section{Results and Discussion}
\begin{figure*}[htbp]
\centering
{\includegraphics[width=\linewidth]{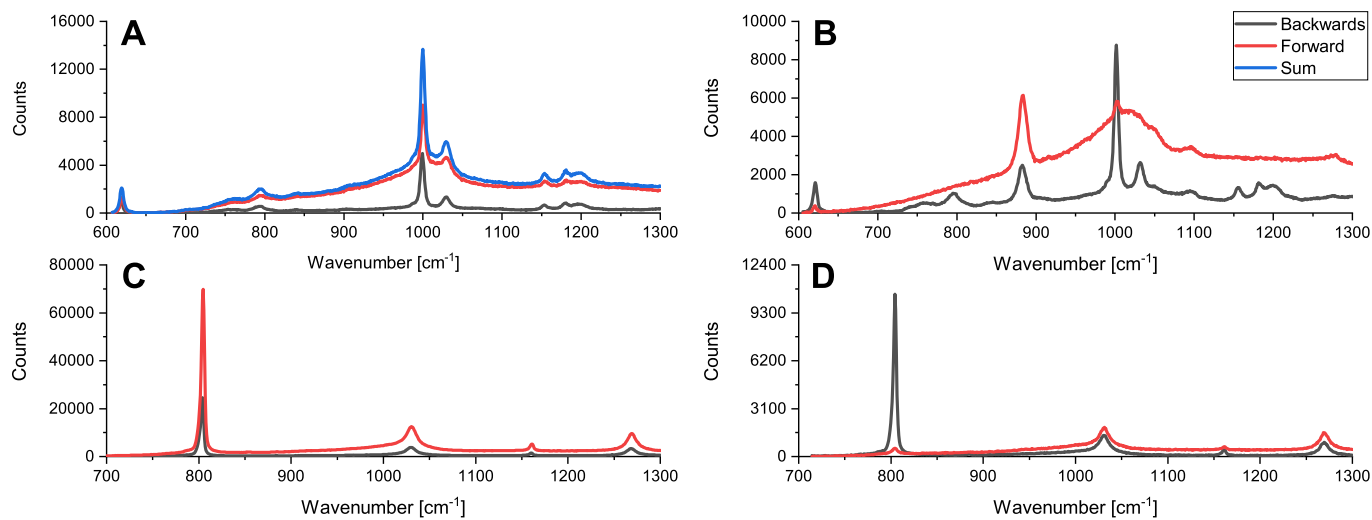}}
\caption{\textbf{A)} Backward (black), forward (red) and sum (blue) spectra of a trapped 3 $\mu m$ polystyrene bead in water. Here we measure with the trapping and condenser objective parfocal to each other. \textbf{B)} A spectrum of a 3 $\mu m$ polystyrene bead trapped in ethanol which adds a peak at 880 $cm^{-1}$. Here the trapping and condenser objectives are defocused by 20 $\mu m$ from each other. \textbf{C)} Spectrum of cyclohexane using pick-off mirror to combine beams measured with the condenser and trapping objective parfocal to each other. \textbf{D)} Using PBS to combine beams and a half-wave plate to rotate the polarization of the forward-scattered beam by 90\textdegree.}
\label{fig:data}
\end{figure*}

To display our setup's increased collection efficiency, we trap 3 $\mu m$ polystyrene beads in MQ water and measure their Raman spectra with the condenser and objective parfocal to each other with the optional confocal pinhole left out. Configured like this, we theoretically collect $\sim$ 60 \% of the total 4$\pi$ solid angle scattered light from the microparticle, thus more than doubling the $28 \%$ collected solely from the back-scattering. This increase in collection efficiency should double the Raman signal if we sum the two detection directions. To verify this experimentally, we compare the intensity of the 1001 $cm^{-1}$ peak between the back-scattered spectrum and the sum and see that the ratio between the two is $2\pm0.1 (n=5)$, close to the expected value of $2$, Fig. \ref{fig:data}. The small deviations in the ratio are probably due to minor alignment differences between samples. These results indicate that even though this method adds complexity to alignment, it is still consistent between samples.

As we can see from the acquired spectrum, the forward-scattered light contains a broad background centered on 1000 $cm^{-1}$, originating from the coverslips that make up the sample chamber, Fig. \ref{fig:data}A. This background is mostly absent from the back-scattered signal due to the confocal pinhole filtering it out. A confocal pinhole gives the back-scattered light a high spatial specificity, meaning it collects light from a small sample volume. Conversely, the forward-scattered light gathers light from a greater sample volume, lacking a confocal pinhole. This ability to collect light from a larger volume results in the forward-scattered light containing more signal from the sample bulk and thus have more background. As a result, our setup effectively gathers light from both the sample bulk and the trapped object simultaneously, adding experimental flexibility to the setup.

To demonstrate this difference in spatial sensitivity between the two detection paths, we measure on 3 $\mu m$ beads suspended in ethanol with the condenser and trapping objective defocused by ~20 $\mu m$ with the optional confocal pinhole left out, Fig. \ref{fig:data}B. In this case, we see that the back-scattered signal looks the same as before except for an added peak at $\sim$880 $cm^{-1}$ corresponding to ethanol. However, the forward-scattered signal has altered. The forward-scattered spectrum now shows a faint polystyrene peak at 1001 $cm^{-1}$ and a prominent ethanol peak. If we compare the ethanol peak ratio between the forward and back-scattered spectrum, we get a ratio of $\sim$2. Thus even when the condenser objective is defocused, it is still twice as efficient as the trapping objective at gathering light from the liquid medium. Also, due to the short depth of focus of the condenser objective, $\sim$0.6 $\mu m$, very little scattered light from the trapped particle gets gathered. Combining these effects, we can see that this setup allows simultaneous measurement of both the bulk sample and the trapped object.

To investigate our setup's capabilities to measure liquid samples, we filled a sample chamber with cyclohexane and measured with trapping, and condenser objective parfocal to each other with the optional confocal pinhole left out, Fig. \ref{fig:data}C. In this case, we see that both objectives gather a clear cyclohexane spectrum. However, if we compare the ratio of the $\sim$800 $cm^{-1}$ peak intensity between the two paths, we see that the forward-scattering is $\sim$2.8 times more efficient. Comparing this to the previous defocused case, we now collect $\sim$40 \% more signal. The cause of this increase is likely twofold. First, defocusing separates the excitation laser's and condenser's focal points, lowering the collection efficiency. Second, in the previous case, a particle was trapped, taking up most of the laser's focal volume, hindering it from exciting the liquid. This result indicates that the setup is beneficial when measuring a purely liquid sample where Raman scattering often is weak.

To further display the setup's flexibility, we show how one can add polarization sensitivity to it. We do this by switching out the D-shaped pick off mirror in the beam combiner with a PBS. In this configuration, the PBS acts as a beam combiner and analyzer to selectively combine the polarized and depolarized Raman bands. As a demonstration, we add a half-wave plate to the forward-scattered beam path and use it to rotate the polarization by 90\textdegree and measure the Raman spectrum of cyclohexane, a chemical with a know polarization dependence \cite{cyklohexanePolar}. From this measurement, we can see a strong polarization dependence on the $\sim$800 $cm^{-1}$ peak, as it has a strongly reduced intensity in the forward-scattered spectrum, Fig. \ref{fig:data}D. As we use an ordinary non-achromatic wave-plate and do not know the two beam paths' collection efficiency, we do not get the polarization dependence in absolute values. However, as proof of concept and to measure relative changes in the polarization dependence of the Raman spectrum of a sample, this method provides a fast and straightforward single-shot way of measuring polarized Raman.

\begin{figure}[h]
\centering
{\includegraphics[width=\linewidth]{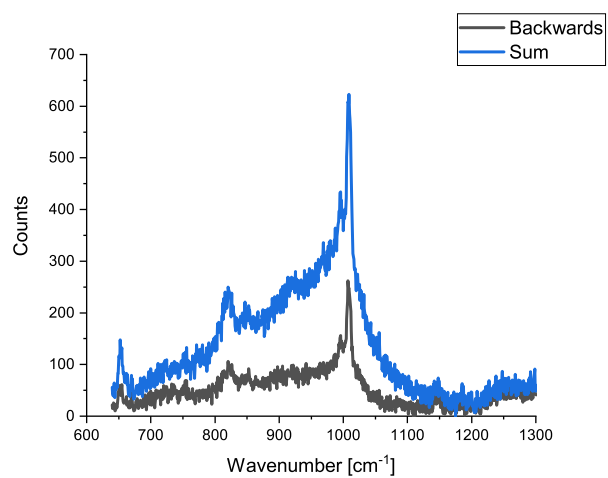}}
\caption{A spectrum of a single trapped \textit{B. Thuringiensis} spore measured using only backward-scattering (black), and a sum of forward and backward-scattering (blue).}
\label{fig:dataSpore}
\setlength\belowcaptionskip{0pt}
\end{figure}

To demonstrate the setup using a biological sample we trapped a \textit{B. Thuringiensis} spore and measured its Raman spectrum, see Fig. \ref{fig:dataSpore}. The data shows a clear \textit{B. Thuringiensis} spectrum with characteristic peaks at 650, 800 and 1016 $cm^{-1}$ even with a modest excitation power of 2.5 $mW$ and integration time of 60 $s$. If we compare the ratio between the 1016 $cm^{-1}$ peaks in the backward-scattered light and the sum of the forward and backward-scattered light, we get a ratio of $\sim$2. It is worth noting that we switched out the standard microscope slides to quartz slides and added the optional confocal pinhole to the forward scattered light to reduce the background noise. The remaining background in the signal probably originates from the immersion oil used to collect the forward-scattered light and could be removed by migrating to a water immersion objective. Thus, we can conclude that this method also doubles the signal intensity even with a confocal pinhole in the forward-scattered light path. As photodamage is an effect that depends linearly on laser power, this means that a doubled collection efficiency allows for a halved photodamage rate which in turn allows for twice as long measurements without extra photodamage. Additionally 

\section{Conclusions}
We demonstrate a simple to implement method that provides a more sensitive and flexible LTRS setup. The added sensitivity allows for decreased laser power and thus reduced photodamage to the sample as well as faster acquisition times. Further, the setup gives the flexibility to measure with the pinpoint spatial selectivity of confocal Raman and the bulk sensitivity of non-confocal Raman simultaneously. Lastly, the setup can measure relative changes in a sample's polarized Raman spectrum in a single shot. We envision the setup being used for faster, less invasive, and more flexible chemical fingerprinting of biological particles.

\section{Funding}
This work was supported by the Swedish Research Council (2019-04016).

\section{Disclosures}
\medskip
\noindent\textbf{Disclosures.} The authors declare no conflicts of interest.

\bibliography{references}


\ifthenelse{\equal{\journalref}{aop}}{%
\section*{Author Biographies}
\begingroup
\setlength\intextsep{0pt}
\begin{minipage}[t][6.3cm][t]{1.0\textwidth} 
  \begin{wrapfigure}{L}{0.25\textwidth}
    \includegraphics[width=0.25\textwidth]{john_smith.eps}
  \end{wrapfigure}
  \noindent
  {\bfseries John Smith} received his BSc (Mathematics) in 2000 from The University of Maryland. His research interests include lasers and optics.
\end{minipage}
\begin{minipage}{1.0\textwidth}
  \begin{wrapfigure}{L}{0.25\textwidth}
    \includegraphics[width=0.25\textwidth]{alice_smith.eps}
  \end{wrapfigure}
  \noindent
  {\bfseries Alice Smith} also received her BSc (Mathematics) in 2000 from The University of Maryland. Her research interests also include lasers and optics.
\end{minipage}
\endgroup
}{}
\end{document}